# Fly ash composites: A review

## Mainak Saha

**Abstract**

The present decade has witnessed numerous investigations focussed on the determination of the mechanical properties of Fly ash composites. These composites have attracted attention, especially in the form of reinforcements owing to their excellent tensile and compressive properties coupled with impact and hardness response. A number of techniques viz. in-situ deposition, hand-up lay and compo-casting techniques have been reported to fabricate these composites. However, in the context of these composites, a systematic structure-property correlation has not been established till date. The present review is aimed at highlighting the current state of research in the avenue of Fly ash composites from two different viewpoints, viz. (i) fabrication technique and (ii) mechanical properties of the fabricated composites. Moreover, the necessity of establishing systematic structure-property correlation in these materials has also been briefly discussed from the author's viewpoint.

**Keywords:** Fly ash composites, stir casting, hand lay-up, mechanical properties, correlative microstructural characterisation.

## 1. Introduction

Composites find a number of engineering applications owing to the enhancement of physical and chemical properties obtained by the combination of two or more materials of which one is a matrix and the other acts as a reinforcement [1]. A number of matrices have been reported, common examples of which include polymer [2], ceramic [3], metals [4], and non-metals [1]. Common examples of reinforcements used in composites are silicon carbide (SiC), boron (B), alumina ($Al_2O_3$), fibers, particulates, graphite, rice husk, and fly ash [5]. Composites are very strong in specific strength, specific modulus, damping capacity, and excellent wear resistance [5]–[8]. Zhang et al. [9] have reported that the mechanical properties of metal matrix composites (MMCs) may be improved by the addition of high strength and high modulus particles like $Al_2O_3$, TiC, $TiB_2$, SiC, etc. as reinforcements.

Two types of fly ash particles have been reported, namely (i) cenosphere-hollow particles (density< 1g/cc) and (ii) precipitator-spherical particles (density range: 2–2.5 g/cc). The former is used in the preparation of ultra-light composites due to their low density when compared to the density of metal matrix whereas, the latter is used for the enhancement of properties which include density, wear resistance, stiffness and strength [10], [11]. Fly ash is a major industrial waste from thermal power plants and is highly abundant in nature owing to its low cost coupled with low density [6], [12]–[14]. **Table. 1** shows chemical composition of fly ash particles (in wt.%) [14]. **Fig. 1** shows Scanning Electron micrograph of fly ash particles [14]. **Table. 2** shows chemical composition (in wt.%) of A356 composite [14]. **Table. 3** shows the Si content (in wt.%) of A356 alloy and that of A356 based C6S, C12S and C12AR composites. Prabu et al. [4] have reported that Fly ash may be used as a potential reinforcement for the fabrication of low-cost composite with enhanced physical and mechanical properties for use in a number of industrial applications. Correlative microscopy, on the other hand, has emerged as a powerful tool for in-depth characterisation of a number of different microstructural features across varying length scales, both structurally and chemically [15], [16]. In other words, the aforementioned tool has the capability to provide both structural and chemical information from the same region in a given microstructure [17]. The present review highlights some of the techniques commonly used for fabricating Fly ash composites. This has been followed by a discussion on the mechanical properties (compressive, tensile and hardness) of these composites. In addition, the importance of correlative microstructural characterisation in terms of establishing a systematic structure-property correlation in these materials has also been discussed from the author's point of view.

**Table. 1** Chemical composition (in wt.) of fly ash particles [14].

| Compound | Composition (wt.%) |
|---|---|
| $SiO_2$ | 64.80 |
| $Al_2O_3$ | 24.01 |
| $Fe_2O_3$ | 5.23 |
| CaO | 2.76 |
| MgO | 0.90 |
| $TiO_2$ | 0.50 |
| Loss on ignition (LOI) | 0.87–1.33 |

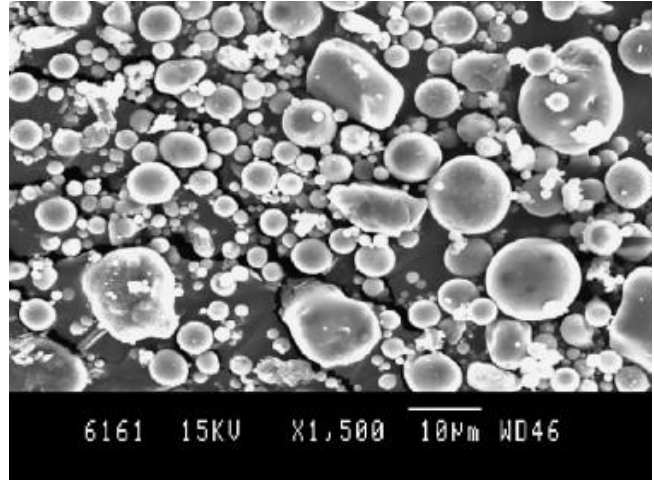

**Fig. 1** SEM image of fly ash particles [14].

**Table. 2** Chemical composition (in wt.%) of A356 alloy [14].

| Element | Composition (wt.%) |
|---------|--------------------|
| Si | 7.083 |
| Mg | 0.409 |
| Cu | 0.059 |
| Ti | 0.132 |
| Mn | 0.003 |
| Cr | 0.012 |
| Al | Balance |

**Table. 3** Si content (in wt.%) of A356 alloy and C6S, C12S, and C12AR composite [14].

| Material | Si (in wt.%) |
|----------|--------------|
| A356Al | 7.08 |
| C6(S) | 8.33 |
| C12(S) | 15.83 |
| C12(AR) | 15.31 |

2. **Fabrication of Fly ash composites**

A number of different techniques have been reported for the fabrication of Fly ash composites. These commonly include stir casting and hand lay-up techniques.

*2.1 Stir casting*

In this method, the dispersed phase is mixed with molten metal matrix by means of mechanical stirring [18]. Large sized composites can be manufactured in a highly economical way [18]. A number of factors such as reinforcement material distribution, wettability, porosity of cast composites and chemical reaction between matrix and reinforcement material are the main process parameters during this process [8].
Fabrication of MMC (by stir casting) using Al356 alloy as a matrix along with the reinforcement of size < 100 μm has been reported by Kulkarni et al. [19]. In addition, fabrication of MMC through combination of (2, 4, 6, and 8) wt% fly ash, (2 and 6) wt% E-glass, and Al6061 composite has been reported [19]. Preparation of three types of Al6061 composites using fly ash as reinforcement for reinforcement weight fractions of 10, 15, and 20% and particle sizes of 4–25, 45–50, and 78–100 μm has been reported using stir casting technique [8]. Al matrix composites in the form of rod and slab have been fabricated by using hexachloroethane tablets, Al6061, and heat-treated fly ash [18].

*2.2 Hand lay-up*

This technique involves the use of hand to combine the resin and reinforcement on the surface of a mold followed by curing of the final laminate without any treatment [7]. The main advantage of this technique is to fabricate large and complex part in a user-friendly manner [8]. For instance, turbine blades of high strength are manufactured at a low cost using this technique. Using this technique, fabrication of glass fiber reinforced fly ash composites with (0, 5,10, and 15 wt.%) reinforcements has been reported [6]. In addition, fabrication of E-glass/epoxy-based composite slabs has been reported with $Al_2O_3$, magnesium hydroxide, hematite powder [19]. Moreover, different concentrations of fly ash (0, 10, and 15 vol.%) have been used to determine the epoxy, filler materials, and fiber fraction [20].

3. **Density**

Fly ash composites have been reported to possess a lower density as compared to that of $Al_2O_3$ reinforcement [19]. In the context of fly ash based composites, density decreases with increasing fly ash content. On the other hand, density increases with increasing $Al_2O_3$ content for $Al_2O_3$ reinforced composites [19]. Boopathi et al. [21] have studied the densities of Al2024 composites reinforced with SiC and fly ash particles using Archimedes principle. It has been reported that density of Al-SiC, Al-fly ash, and Al-SiC-fly ash composites undergoes a linear decrease with increasing fly ash and SiC contents. This may be attributed to the lower density of fly ash and SiC particles with respect to that of Al (matrix) [21]. **Table. 4** shows the theoretical and measured densities and porosities of A356 alloy and A356 based C6S, C12S, and C12AR composites in both as-cast and extruded conditions. Density of composites was experimentally measured using Archimedes principle [14].

**Table. 4** Densities and porosities of A356 alloy and A356 based C6S, C12S, and C12AR composites [14].

| Material | Theoretical density (g/cc) | Measured density | | Porosity (vol. %) | |
|---|---|---|---|---|---|
| | | As-cast | Extruded | As-cast | Extruded |
| A356 Al | 2.680 | 2.665 | 2.676 | 0.560 | 0.149 |
| C6(S) | 2.645 | 2.616 | 2.640 | 1.096 | 0.189 |
| C12(S) | 2.609 | 2.442 | 2.591 | 6.401 | 0.690 |
| C12(AR) | 2.611 | 2.458 | 2.602 | 5.860 | 0.345 |

## 4. Mechanical properties

*4.1 Tensile properties*

The tensile strength of the fly ash composites has been reported to be increased by the addition of fly ash content and E-glass fiber in Al6061 [18]. While applying tensile load, the presence of high-strength fly ash particles leads to an increase in tensile strength [22]. The increase in ash content in Al7075 alloy leads to an enhancement in the tensile strength of the alloy up to a certain limit beyond which it decreases [23]. **Fig. 2** shows the tensile fracture surface of fly ash reinforced A356 composites [14].

It has also been observed that the tensile strength of fly ash and alumina reinforced LM25 MMC increases with increasing fly ash content [24]. The variation of the tensile strength of polyphenylene oxide composite with fly ash reinforcement has been experimentally investigated [25]. It has been shown that for higher size fly ash content, tensile strength (ultimate tensile strength (UTS)) is independent of fly ash content, whereas for smaller particle sizes, UTS decreases due to the agglomeration for high fly-ash concentration [25].

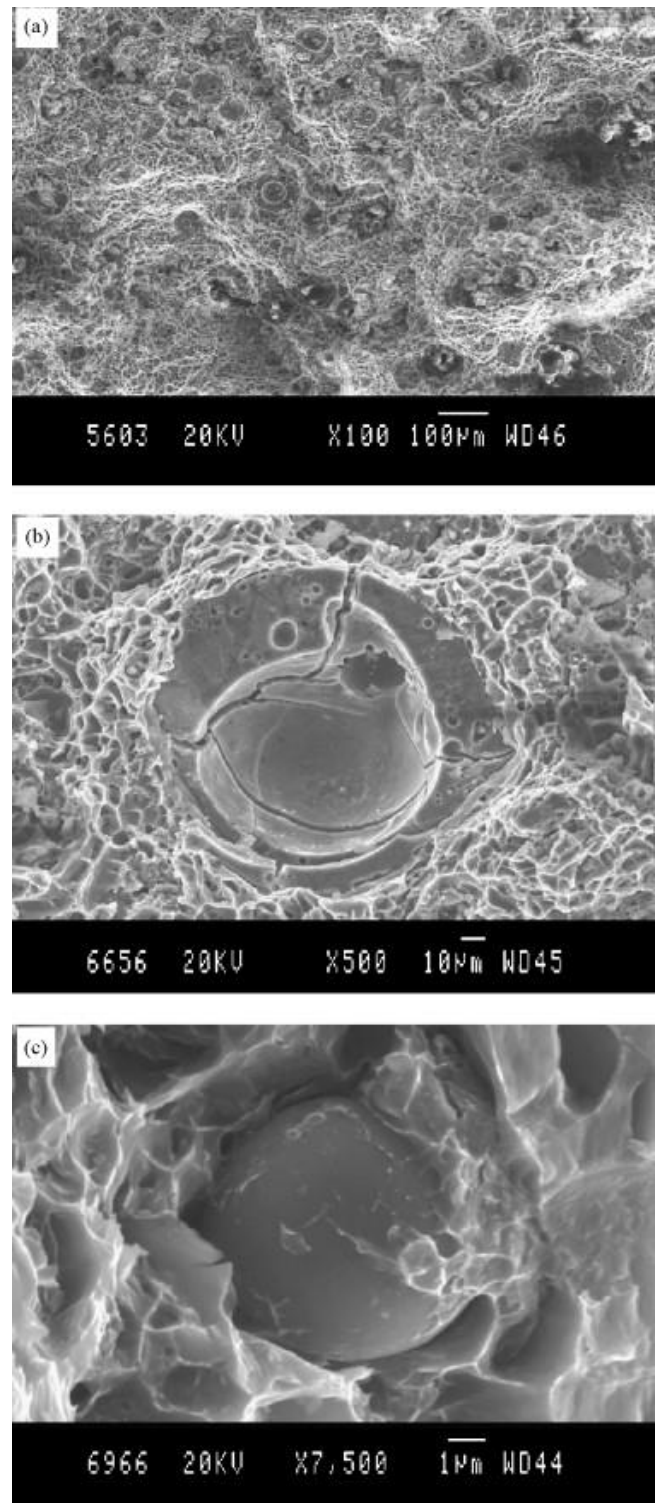

**Fig. 2** SEM image of fracture surfaces (after tensile fracture) of fly ash reinforced A356 composites [14].

*4.2 Compressive properties*

In all types of fly ash reinforced composites, the compressive strength increases with increase in fly ash content. However, composites reinforced with $Al_2O_3$ show higher compressive strength than the one reinforced with fly ash [19]. Besides, reinforced with E-glass fiber and fly ash lead to an increase in the density (of the composite) which enhances the compressive strength of these composites [19]. In the context of Al6061, reinforcement with fly ash particles has been reported to lead to an increase in compressive strength of the composite [8]. The compressive strength of the composite has also been reported to decrease with increasing fly ash particle size . In the context of concrete composite reinforced with fly ash and nano-$SiO_2$, compressive strength is found to be higher than that of concrete composite without the presence of steel fiber reinforcement [9]. **Fig. 3** shows SEM image of fly ash reinforced A356 composite showing fractured particles of fly ash after compression [14].

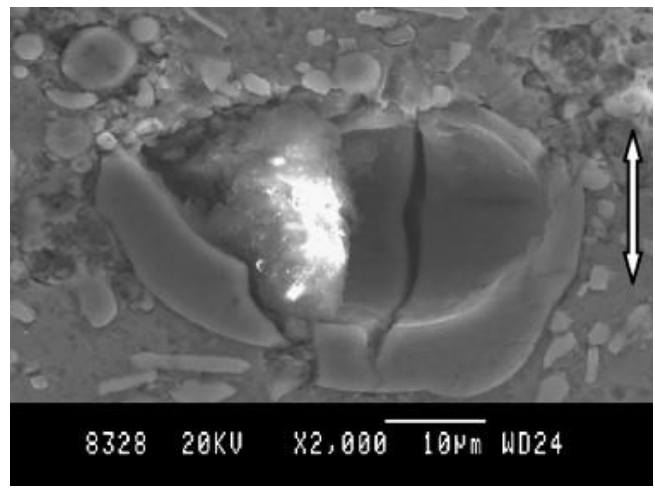

**Fig. 3** SEM image of fly ash reinforced A356 composites showing fractured particles (after compression) of fly ash [14]. The white arrow in the image represents loading direction.

*4.3 Hardness*

In the context of Al6061 composites, there is a conversion from ductile to brittle nature with increasing fly ash content [7]. This leads to an increase in the hardness of these composites [23], [25]. In the case of polymers, resistance to plastic deformation is increased by the addition of fly ash. This leads to an increase in the hardness value. For instance, the 300-nm-size fly ash filler in polymer matrix composites results in higher hardness when compared to other fly ash filler sizes [26]. This has been attributed to a larger phase contact between the filler and matrix phases [27]. The hardness of these composites has also been reported to increase with increasing silicon carbide (SiC) particle (as reinforcement). For the combination of

Al/(10%SiC+10%fly ash), maximum hardness is achieved [28]. **Table. 5** shows Vickers microhardness and Brinell microhardness values of A356 alloy and its composites.

**Table. 5** Vickers and Brinell hardness of A356 alloy and its composites [14].

| Alloy/composite | Vickers hardness number (VHN) | Brinell hardness number (BHN) |
|---|---|---|
| A356 Al | 51.1 ± 1.2 | 48.5 ± 1.0 |
| C6S | 59.5 ± 1.1 | 54.8 ± 0.5 |
| C12S | 55.6 ± 2.5 | 54.2 ± 0.4 |
| C12AR | 54.3 ± 0.8 | 52.4 ± 0.9 |

*4.4 Wear*

In the context of Al MMCs reinforced with fly ash, Sharma et al. [29] have reported that wear resistance of the fabricated composites increases with increasing fly ash content. In addition, composites with high fly ash contents have been reported to result in ~13.6% less wear as compared to that of the composites with low fly ash content [29]. For medium fly ash content (~ 4%) average coefficient of friction was reported to be the lowest (~0.12) whereas, for high fly ash contents (~6%), the average coefficient of friction was reported to be the maximum (~0.161). Based on this work, it was concluded that the amount of fly ash content in the Al matrix must not be > 4%, so as to prevent a high value of the coefficient of friction. Sudarshan and Surappa [11], [14] and Rohatgi et al. [8] have studied the dry sliding wear of fly ash particle reinforced A356 Al composites. It was reported that incorporation of 6 vol. % of fly ash particles into A356 Al alloy leads to a decrease in dry sliding wear rates at low loads (~10 and 20 N) [11]. 12 vol.% of fly ash reinforced Composites reinforced with ~12 vol.% of fly ash were reported to show lower wear rates as compared to that of non-reinforced alloy between 20 and 80 N [8], [11]. For composites with 12 vol. % fly ash, wear rate was observed to decrease with particle size [11]. An increase in the magnitude of friction co-efficient (from ~0.49 to ~0.58) with increasing fly ash content from 6 to 12 vol.%) was reported [8], [14]. Adhesive wear was reported to be dominant in non-reinforced alloy, whereas abrasive wear was prevalent in composites [8], [11]. Moreover, at higher load, subsurface delamination was reported to be the main mechanism for both alloys and composites [8]. **Fig. 4** shows the SEM image of worn surfaces (at various loads) of fly ash reinforced A356, C6S, C12S, and C12AR composites [11]. In the context of Al-10 wt.% fly ash reinforced composite, Desai et al. [5] have reported that high load results in high pressure at the point, of contact subsequently resulting in a high rubbing action [5]. As a result, an increase in the applied load leads to an increase in the wear rate [5]. It is worth mentioning that wear mechanisms involving adhesion are predominant in the base metal, while abrasion with micro-cutting and oxide formation dominate in Al based fly ash composites [5].

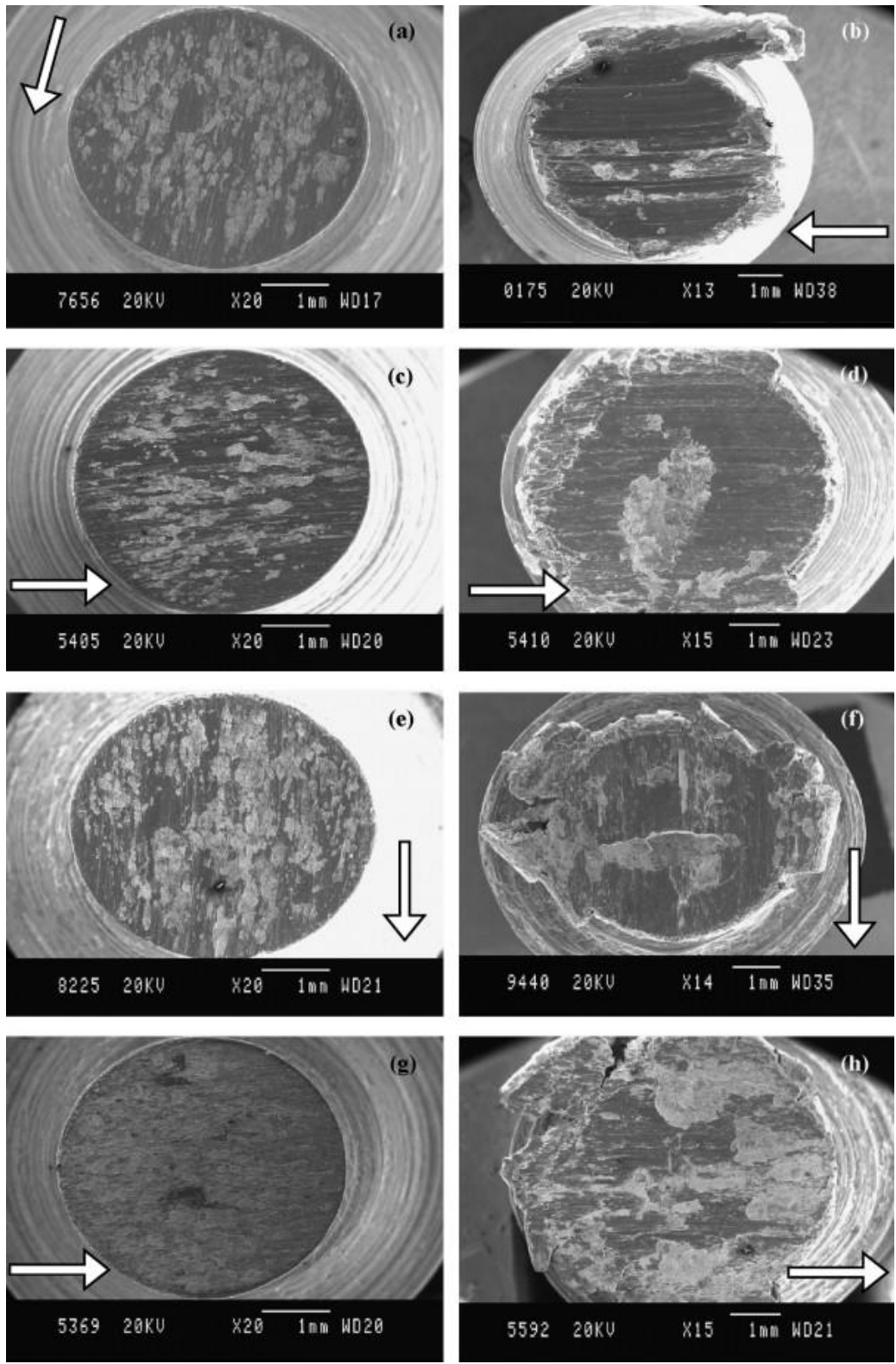

**Fig. 4** SEM image showing worn surface of fly ash reinforced A356 composite at various loads: **(a)** 10 N, and **(b)** 80 N; C6S composite: **(c)** 10 N and **(d)** 20 N; C12S composite: **(e)** 10 N, and **(f)** 80 N; and C12AR composite: **(g)** 10 N and **(h)** 80 N [11]. In parts **(a-h),** AR and S

abbreviate for as-receive and sieved respectively. Besides, C6S represents 6 vol.% of fly ash (as a reinforcement). Similarly, C12S represents 12 vol.% of fly ash (as a reinforcement) [11].

## 5. Future perspectives

The present state of research in the field of fly ash composites is focussed on the fabrication techniques and mechanical properties of these materials. However, a systematic structure-property correlation based investigation is missing. This is because research in the field of these composites, till date, has not been dedicated to: (i) understanding the influence of process parameters (associated with the different fabrication techniques, as discussed in section **2**) on the microstructure evolution, and (ii) the understanding of microstructural defects and their associated influence on the mechanical properties of these materials. This may be attributed to the fact that at present, most research groups involved with fly ash composite research are not primarily focussed on metallurgical research.

The mechanical properties of composites are highly sensitive towards microstructural defects, especially developed during fabrication [30]. Moreover, these properties are anisotropic along longitudinal and transverse section of these composites [31], [32]. In the last 10 years, correlative microscopy has emerged as a powerful tool for a structural cum chemical characterisation from the same region in a given microstructure [1], [33]. This has been largely utilised for understanding the influence of microstructural defects on the mechanical properties of metallic materials [34]. However, till date, this methodology has not been utilised towards establishing systematic structure-property correlation in these fly ash composites. The challenge primarily lies in terms of the expertise and cost associated with operating Focussed Ion Beam (FIB)-based sample preparation technique for preparing samples for correlative microstructural analysis across different length scales and the reproducibility of the results obtained. The other challenge lies in understanding the complexity of crystal structures associated with these composites. Understanding crystal structures may be considered as the first step towards addressing microstructural defects associated with processing of these composites. Hence, the aforementioned methodology shows extensive potential for future research in the avenue of fly ash research.

## 6. Conclusions

The present need of the hour is to design composites with high specific strength. Fly ash particles have shown tremendously high potential as reinforcements with lightweight Al alloys leading to enhanced mechanical properties. More specifically, hardness, tensile and compressive strength of these composites increase with increasing fly ash content. On the other hand, density decreases with increasing fly ash content. However, extensive microstructure analysis aimed at systematic structure-property correlation in these materials is necessary before using these materials for large-scale industrial applications. Correlative microscopy, as a characterisation methodology, shows a high potential towards extensive structural cum chemical characterisation of these materials. This methodology may be used to obtain an in-depth understanding of the microstructure evolution in these materials.

**Acknowledgement**

MS would like to thank the research scholars in the Department of Metallurgical and Materials Engineering, NIT Durgapur, for detailed discussions during the scripting of the review paper.